\documentclass{elsart}

\usepackage{natbib}

 \usepackage{graphics}

\begin{document}

\begin{frontmatter}



\title{Properties of SN-host galaxies}


\author{F. Combes}

\address{Observatoire de Paris, LERMA, 61 Av. de l'Observatoire,
F-75014, Paris, France}
\ead{francoise.combes@obspm.fr}

\begin{abstract}
It is of prime importance to recognize evolution and
extinction effects in supernovae results as a function 
of redshift, for SN Ia to be considered as distance indicators.
This review surveys all observational data searching
for an evolution and/or extinction, according to host morphology.
For instance, it has been observed that
high-z SNe Ia have bluer colours than the local ones: although this 
goes against extinction to explain why SN are dimmer
with redshift until z $\sim$ 1, supporting a decelerating universe,
it also demonstrates intrinsic evolution effects.
\\
SNe Ia could evolve because the age and metallicity of their
progenitors evolve.
The main parameter is carbon abundance. Smaller C leads to a dimmer SN Ia
and also less scatter on peak brightness,
as it is the case in elliptical galaxy today.
Age of the progenitor is an important factor: young populations lead
to brighter SNe Ia, as in spiral galaxies, and a spread in ages
lead to a larger scatter, explaining the observed lower scatter at high z.
\\
Selection biases also play a role, like the Malmquist bias;
high-z SNe Ia are found at larger distance from their host center: 
there is more obscuration
in the center, and also detection is easier with no contamination 
from the center.  This might be one of the reason
why less obscuration has been found for SNe Ia at high z.
\\
There is clearly a sample evolution with z:
currently only the less bright SNe Ia are
detected at high z, with less scatter.
The brightest objects have a slowly declining light-curve, and at high z,
no slow decline has been observed. This may be interpreted as an age effect,
high-z SN having younger progenitors.
\end{abstract}


\end{frontmatter}

\section{Introduction: SN Ia as ``calibrated'' candles}
Type Ia supernovae are now used as fundamental probes
of the cosmological parameters, based on a tight
empirical relation between their peak luminosity and
the width of their light-curve (Riess et al. 1996, Perlmutter et al. 1997).
 It has been recognized early than SN Ia were not "standard"
candles, since important variations of their peak luminosities
are observed, as a function of metallicity, age,
environment, morphological type of the supernovae hosts.
 But most of these variations can be calibrated, and corrected away,
if the correlation between the peak luminosity and the rate of
decline is taken into account (Phillips 1993): the scatter in
distances is then much reduced.
The problem arises with evolution, since these effects are
not known a priori, and could mimick a cosmological constant.
\\
In particular the presence and nature of dust could vary
with redshift, or the rate of supernovae explosions and
their nature could vary systematically.
A striking example of evolution of supernovae with redshift is
the observation that
high-z SN Ia have bluer colours than local ones (cf Fig. 1, 
Leibundgut 2001).
This result is surprising, since it goes against extinction and reddening.
\begin{figure}[h]
\centerline{\resizebox{9cm}{!}{
{\includegraphics{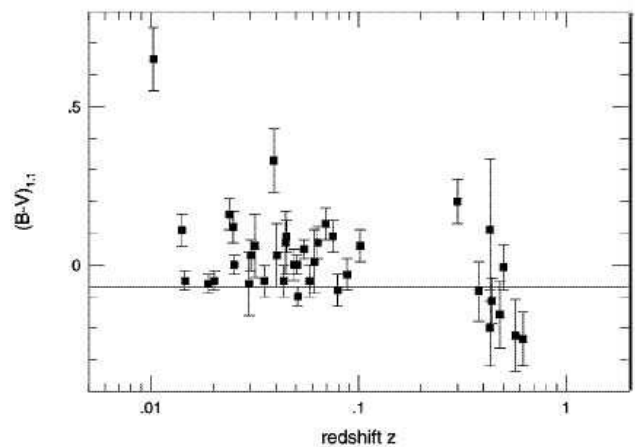}} 
}}
\caption{ The observed (B-V) color of SNe Ia versus redshift. The horizontal
line shows the intrinsic color for the nearby SNe Ia, 
from Leibundgut (2001).}
\label{fig1}
\end{figure}
\section{Evolution of SN Ia, corrections and consequences}
 To better understand the origin of these evolutions,
we must know more on the progenitors of these SNe,
to predict age or metallicity effects.
\subsection{Progenitors of type Ia supernovae}
Type Ia supernovae are likely to be thermonuclear explosion (by fusion
of carbon and oxygen) of
white dwarfs having approached the Chandrasekhar mass
of 1.39 M$_\odot$ (cf the review by Hillebrandt \& Niemeyer 2000).
Their principal characteristic constraining models is their absence
of hydrogen in their spectrum.
They correspond to the total explosion of a
Carbon-Oxygen white dwarf (C+O WD). They must accrete
mass through binary evolution (since these WD are not massive enough at
the beginning), and there are two possibilities:
either two C+O white dwarfs merge (double-degenerate model), in a time-scale of 
a few 100 Myr, or only one C+O WD is accreting (single degenerate), 
with a longer time-scale,
a few Gyrs, the most likely scenario.  The merger scenario
is not supported by the homogeneity of observed quantities, since
the obtention of the critical mass by fusion introduces more freedom
in initial parameters.  The range of luminosities is assumed to be connected
to the amount of radioactive $^{56}$Ni produced in the explosion, decaying to 
 $^{56}$Co and $^{56}$Fe. The amount of nickel available
cannot vary by much more than a factor 2.
The merger scenario is supposed to lead to a collapse supernova instead.
\\
Recently, hydrogen (broad H$\alpha$ line) has been discovered in a type Ia supernova
(Hamuy et al 2003), which has been considered to support the single-degenerate
progenitor. However, it could be interpreted as a double-degenerate case also
(Livio \& Riess 2003), and in any case is too rare to clarify definitively
the issue.
Some of the observed variations could be due to the various time-scales
and ages of the progenitors: the
most luminous SN Ia are found in spirals, the dimmest in Ellipticals.
The rate of SN Ia has been estimated (Pain et al. 1996, 2002), and could
bring insight into the problem: 6\% of the stars between 3-9 M$_\odot$ 
experience such an explosion.
\subsection{Effects of age and metallicity}
Smaller Carbon abundance of progenitors leads to dimmer type Ia supernovae
(Umeda et al 1999). The C fraction varies between 0.36-0.5 just before the explosion.
The mass of the progenitors are smaller at lower metallicity, and the
model predicts dimmer SN at high z, with lower scatter.
Timmes et al. (2003) confirm that the peak luminosity should vary linearly
with metallicity Z, and this could already explain a scatter of 25\% for
a region like the solar neighborhood, where Z varies between 0.3 and 3 solar.
However, observations do not confirm this strong predicted influence of metallicity: 
in  elliptical and S0 galaxies, where metallicity varies with radius,
no significant radial gradients of SNe Ia peak absolute magnitudes or decay rates
have been detected (Ivanov et al. 2000). The variations 
in brightness and light curve width of supernovae reveal high values, that are attributed
more to age. Once the light-curve shape method is applied, there does
not remain any radial gradient of colours. This supports the validity 
of the empirical calibrations.
\\
The effect of age could be partly due to metallicity (stars being
less metallic in the past), but also due to different explosions
scenarios. At high redshift the progenitors were obviously younger,
which favors the short-time scale double-degenerate scenario. However
this explosion model leads to brighter SNe, which is not observed.
Brightest objects have a slowly declining light curve. At high z
no slow decline are observed, as if the brightest SN are not seen,
(contrary to the Malmquist bias).
\begin{figure}
\centerline{\resizebox{12cm}{!}{
{\includegraphics{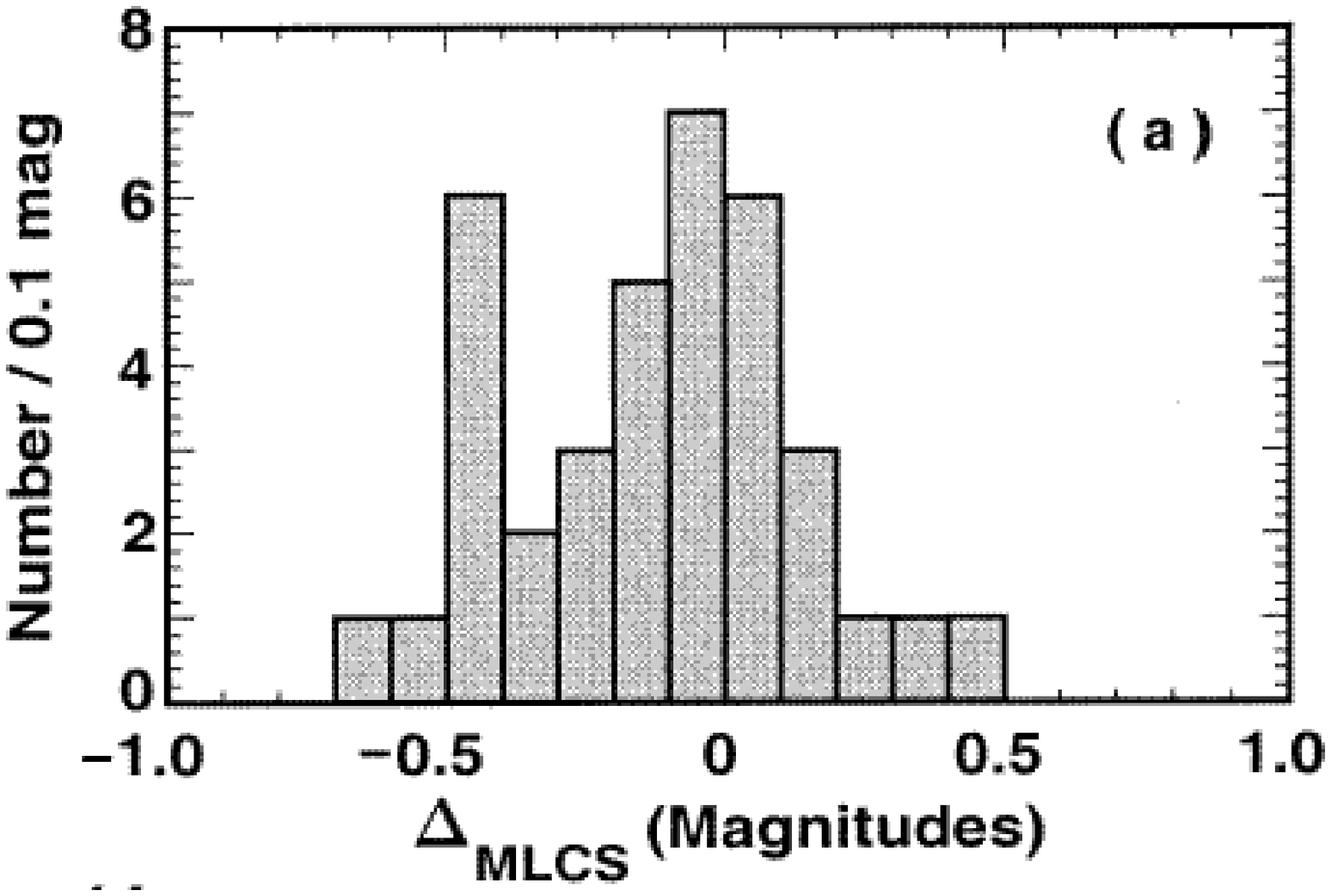}} 
{\includegraphics{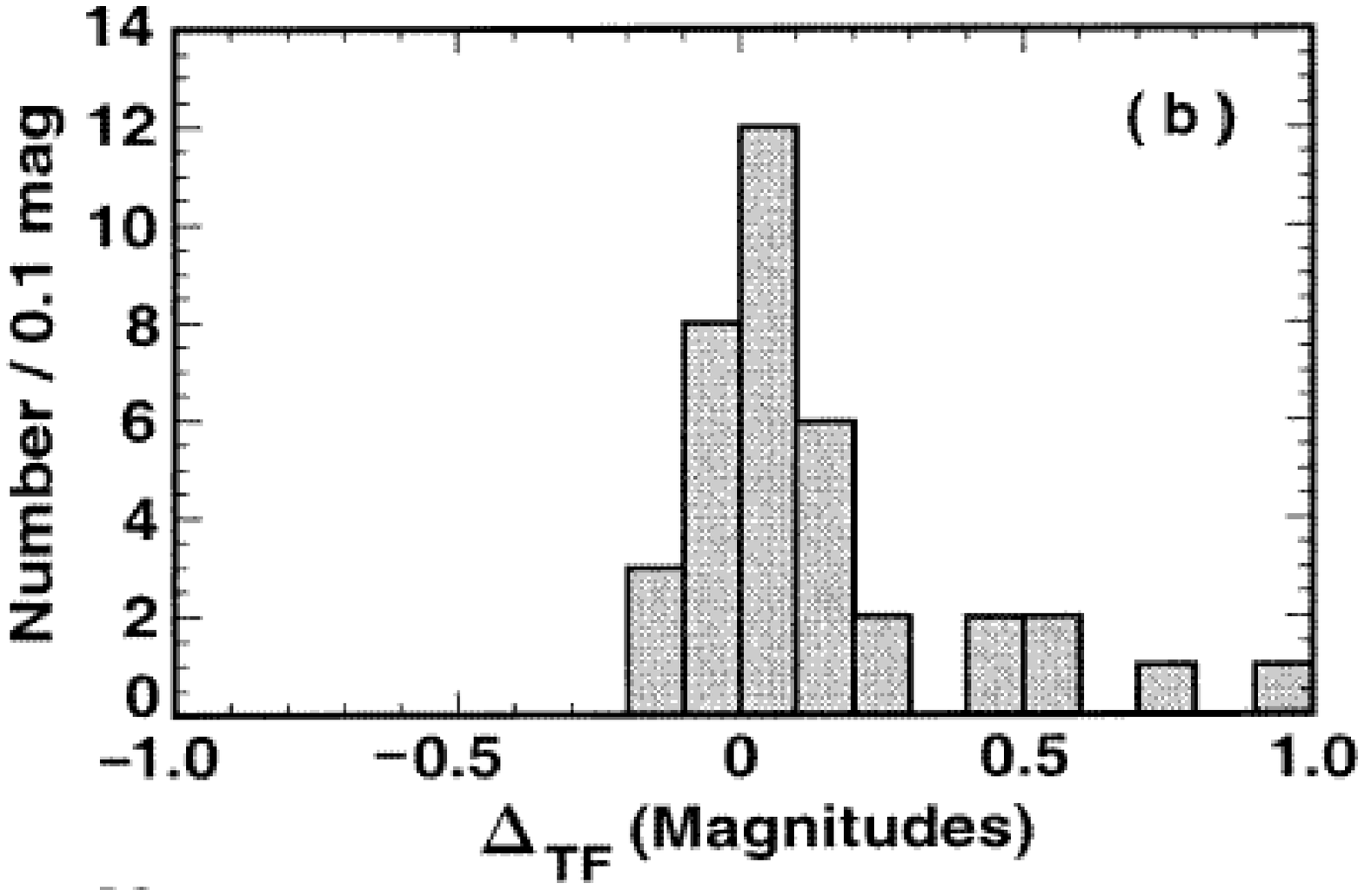}} 
}}
\centerline{\resizebox{6cm}{!}{
{\includegraphics{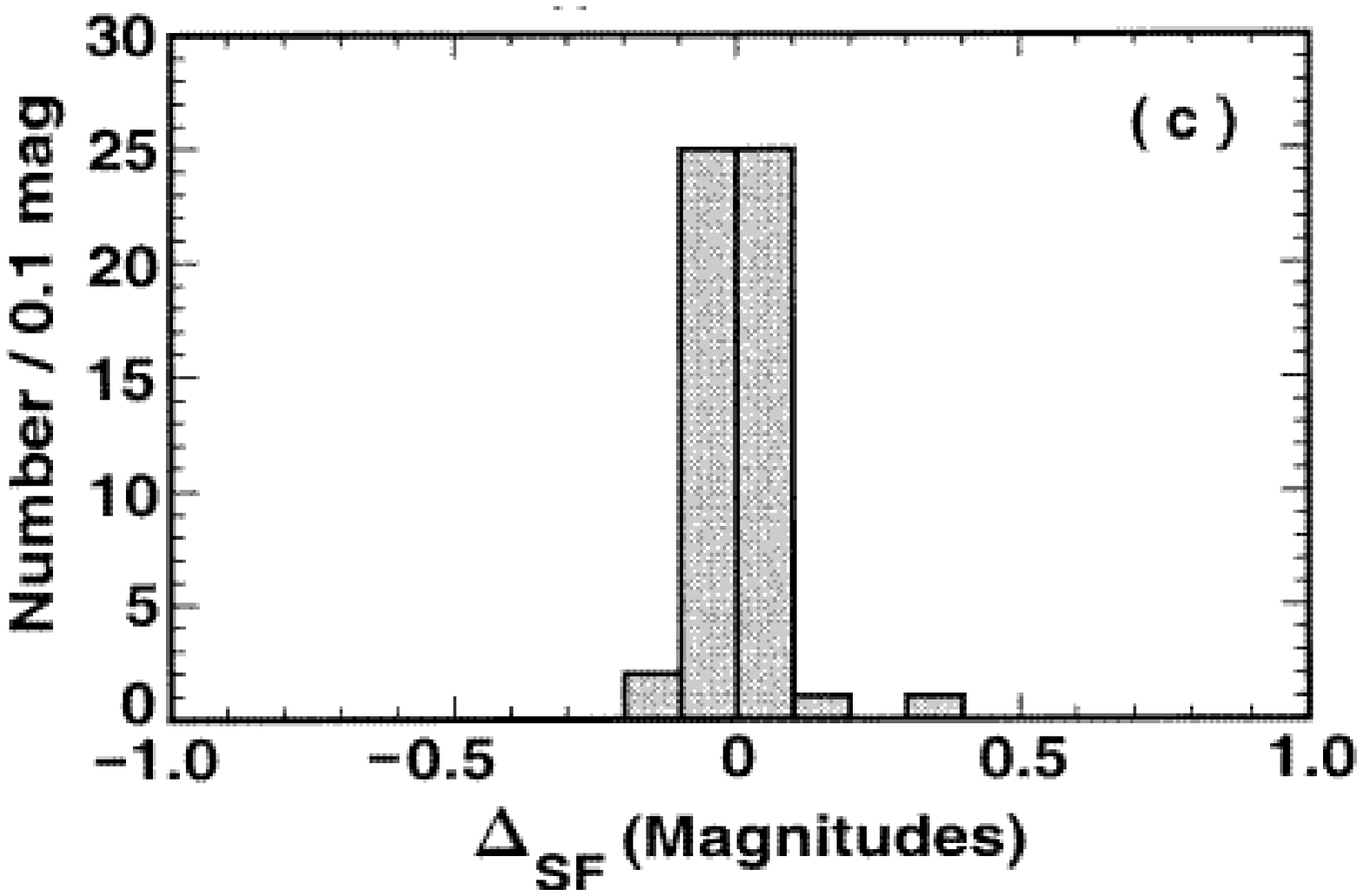}} 
}}
\caption{ Histograms of the corrections to the absolute magnitudes
of the observed SNe Ia, for the three methods used:
MLCS (Multicolor Light-Curve Shape), TF (Template Fitting), and
SF (Strech Factor), from Drell et al. (2000).}
\label{fig2}
\end{figure}
\subsection{Corrections toward standardisation}
The relation between the peak luminosity of SNe Ia and their initial
rate of decline, is different in the various bands: the decline is faster
in the B-band, than in V and I, and the colours at maximum depends
on the decline rate (Phillips 1993); the fastest
declining light curves (the less bright objects)
correspond to the reddest events.  All these characteristics are used
to reduce the range of peak luminosities, and to approach the standard
candle regime. Various methods have been used by the two main teams,
the Supernova Cosmology Project (SCP, Perlmutter et al 1997), and the
High-Redshift Supernova Search team (HZT, Riess et al 1998). The methods
are (see Fig. 2):
\\
-- MLCS: Multicolor Light-Curve Shape
(HZT), calibrated on nearby SNe Ia in the local universe (z $<$ 0.15),
uses the color characteristics to estimate corrections for extinction and reddening
due to material in the host galaxies;
\\
-- TF: Template Fit  (HZT) uses several template curves, from peak to 
15 days afterwards, to fit a particular high-z SN Ia;
\\
-- SF: Stretch Factor (SCP): the shift
in the peak magnitude is proportional to how much
the light-curve width must be stretched to fit a standard curve.
\\
What is striking in the comparison of the various methods in 
Fig. 2 (made by Drell et al. 2000) is that the amount of corrections
are quite different, and in particular the SF method corrections are
very small, meaning that these SNe Ia are almost standard candles.
The samples are different, although 14 objects are in common for the 3 methods.
The study by Drell et al. (2000) did not reveal any
systematic  differences in distances between the two methods
MLCS and TF as a function of redshift (only a scatter of about 0.4 mag), 
but a little dependence on absolute magnitude, for the high-z sample.
\\
Drell et al. (2000) made interesting simulation  
of cosmic evolution of SNe, as for instance their peak luminosity varying
slightly with (1+z) linearly, and showed clearly how the determination of cosmological
parameters and cosmic evolution is degenerate. With the number of high-z SNe Ia
observed up to now, observing a larger number will not raise the degeneracy, but
a better understanding of the explosion process is more needed.
\section{Extinction}
The possibility that intergalactic dust, distributed uniformly between galaxies,
produce the extinction able to account for the dimming of high-z SNe Ia, has been
investigated by  Aguirre (1999) and Aguirre \& Haiman (2000). The exctinction
must not be accompanied by reddenning, which is not observed (this must be
gray dust). This hypothesis is compatible with the data, 
given that a large amount of gas is observed in the intergalactic space
(Ly$\alpha$ absorbants), and that the metallicity of the gas appears high
(0.1 to 0.3 solar), a significant fraction of metals can be intergalactic. 
 The grains have to be of large size ($>$ 0.1$\mu$) for the ``gray'' requirement.
Very small grains provide most of the reddening but less than half of the 
opacity for optical extinction. This gray dust would however 
absorb the cosmic UV-optical background and by re-emission,
contribute significantly (75\%) to the cosmic infrared background (CIB).
There are no measurable distortions of the CMB predicted, but
the dust FIR background is only marginally compatible with the data.
Also, this component is not detected by X-ray scattering
(Paerels et al 2002).  The extinction could as well be provided by 
intervening galaxies, but this hypothesis appears
unlikely also (Goobar et al 2002).
\\
Extinction from host galaxies appears more likely,
although high-z SNe Ia are not reddened,
and apparently not obscured, according
to their correction factors, which is difficult to explain, except
through selection effects and Malmquist bias (Farrah et al. 2002).
The proportion of morphological types for the hosts do not show evolution:
70\% are found in spiral galaxies and 30\% in ellipticals, as
for local events.
\subsection{Influence of distance from center}
The radial distribution of SNe in their host galaxies is  observed to be similar
at low and high z, at least for the samples observed with CCD, 
but not with the samples discovered photographically
(Shaw effect (cf Shaw 1979),
where the center of galaxies is undersampled in SNe, Howell et al. 2000).
SNe in the outer regions of galaxies are dimmer than those in the central regions 
by 0.3 mag  (Riess et al. 1999). This could explain the anti-Malmquist bias
observed for high-z SNe. Another concording  feature is that
SNe in early-type galaxies tend to be dimmer, and large-distance
SNe are in general in early-types (Hamuy et al. 1996).
Physical explanations for the dependence of peak luminosity
with radius might be searched in the 
metallicity gradient of the galaxy host, or in the possible
different progenitors between disk and bulge (age effect).
\subsection{Influence of dust and host-type}
Sullivan et al (SCP, 2003) have recently studied in details coherent
samples of SNe Ia, and their host morphologies. They built a Hubble
diagram (Fig. 3) according to the host types, omitting outliers 
(too large stretch factors, highly reddened objects).
The classification of the hosts is made from colours, HST images and spectral types
when available, the evolution of colours with redshifts being compared to
stellar population models (Poggianti 1997).
When stretch factors (SF) are compared for all types, early-types SF are more
scattered. A striking feature is that SF are less scattered for high-redshift
objects.  This might be due to the age effect: at low redshift, 
stellar populations have a larger range of ages than at high z.
\\
The Hubble diagram shows more scatter for late-type hosts, 
which might be attributed to dust extinction, since
there is more gas in late-type galaxies.
There is also more scatter  
with SNe discovered in the central parts
of the galaxies, which confirms that the scatter 
might be due to dust. There is no or little evolution
of the amount of dust extinction with redshift.
The estimated average extinction suffered by SNe Ia is small
(A$_B$ $\sim$ 0.3, but may be $>$1 when all SNe are considered,
including outliers). This is consistent with what is expected
from galaxies, where the mean exctinction is weighted by
a small number of objects highly obscured. Selection effects
eliminate the more obscured SNe. 
\begin{figure}
\centerline{\resizebox{9cm}{!}{
{\includegraphics{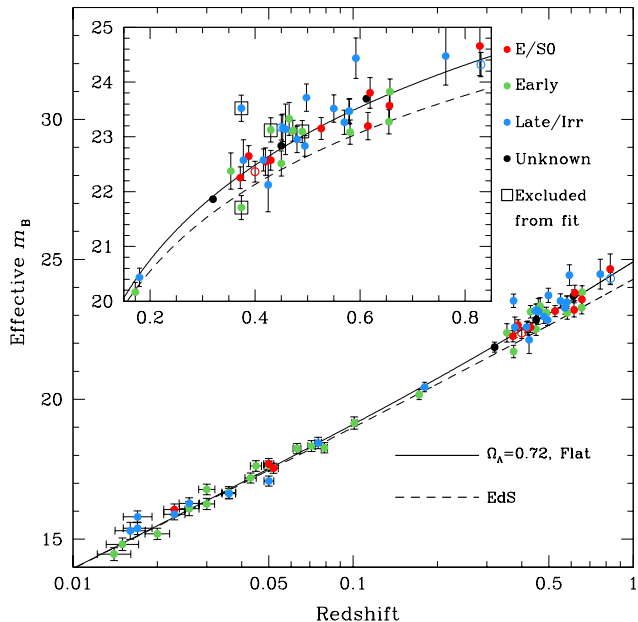}}
}} 
\caption{ The Hubble diagram for the SF-corrected SNe Ia, plotted with
different symbols according to the host-galaxy type. The inset
details the high-z part of the diagram.
The best fit flat cosmological model ($\Omega_m$ =0.28, 
$\Lambda$=0.72), is the full curve, from Sullivan et al. (2003).}
\label{fig3}
\end{figure}
\section{Supernova rates}
Since the peak luminosity of SNe varies with galaxy hosts,
it is interesting to study the SNe rates as a function
of galaxy types, and evolution of types with redshift.
As for core-collapse supernovae (CCSN) and in a 
less measure for SNe Ia, it is legitimate to expect a larger
SNe rate for starbursting galaxies. Starburst galaxies are
dusty, and this rate has been studied in 
the infrared (Mattila \& Meikle 2001;
Mannucci et al. 2003). The rate of supernovae in starburst
galaxies is found to be about
10 times more than in quiescent galaxies.  But it is still 3-10 times
less than expected from the star formation rate (SFR) estimated with FIR
emission, which means that many SNe are still obscured (Av $>$ 30).
Alternatively, the FIR tracer of star formation might not 
follow closely the SN rate, as the latter is found also relatively too
large in low-mass galaxies (Melchior et al. 2003).
Other possibilities is that starburst and quiescent galaxies have
different IMF, leading to a different SN-rate/SFR ratio.
\\
The type Ia SN rate shows also a sharp increase toward galaxies 
with higher activity of star formation,
suggesting that a significant fraction of type Ia SNe 
are associated with young stellar populations
(della Valle \& Livio 1994; Mannucci et al. 2003).  
This is important, in view
of the large increase of star formation activity with redshift.
\\
It is well known also that the supernova rates is
strongly correlated with Hubble types 
(Branch \& van den Bergh 1993; Hamuy et al. 2000).
Late-type systems are more prolific in SNe Ia,
but selection effects are severe, and it is difficult
to disentangle effects of age and metallicity.
\section{Galaxy types and star formation history}
Large statistics are needed to better understand the
evolution of star formation as a function of galaxy
types, and in consequence the SN rate as a function of host
type. Large surveys of galaxies are currently carried out,
and the first results on 10$^5$ galaxies in 
the SLOAN Digital Sky Survey (SDSS)
give already such statistics (Kauffmann et al 2003a,b).
Two stellar absorption line indices (the strength of the
4000\AA i\, break D$_n$(4000), and the Balmer 
absorption line index H$\delta_A$) with the help of models,
yield the amount of dust extinction, the stellar masses and
the fraction of mass involved in bursts in the last 1-2 Gyr.
\begin{figure}[h]
\centerline{\resizebox{5cm}{!}{
{\includegraphics{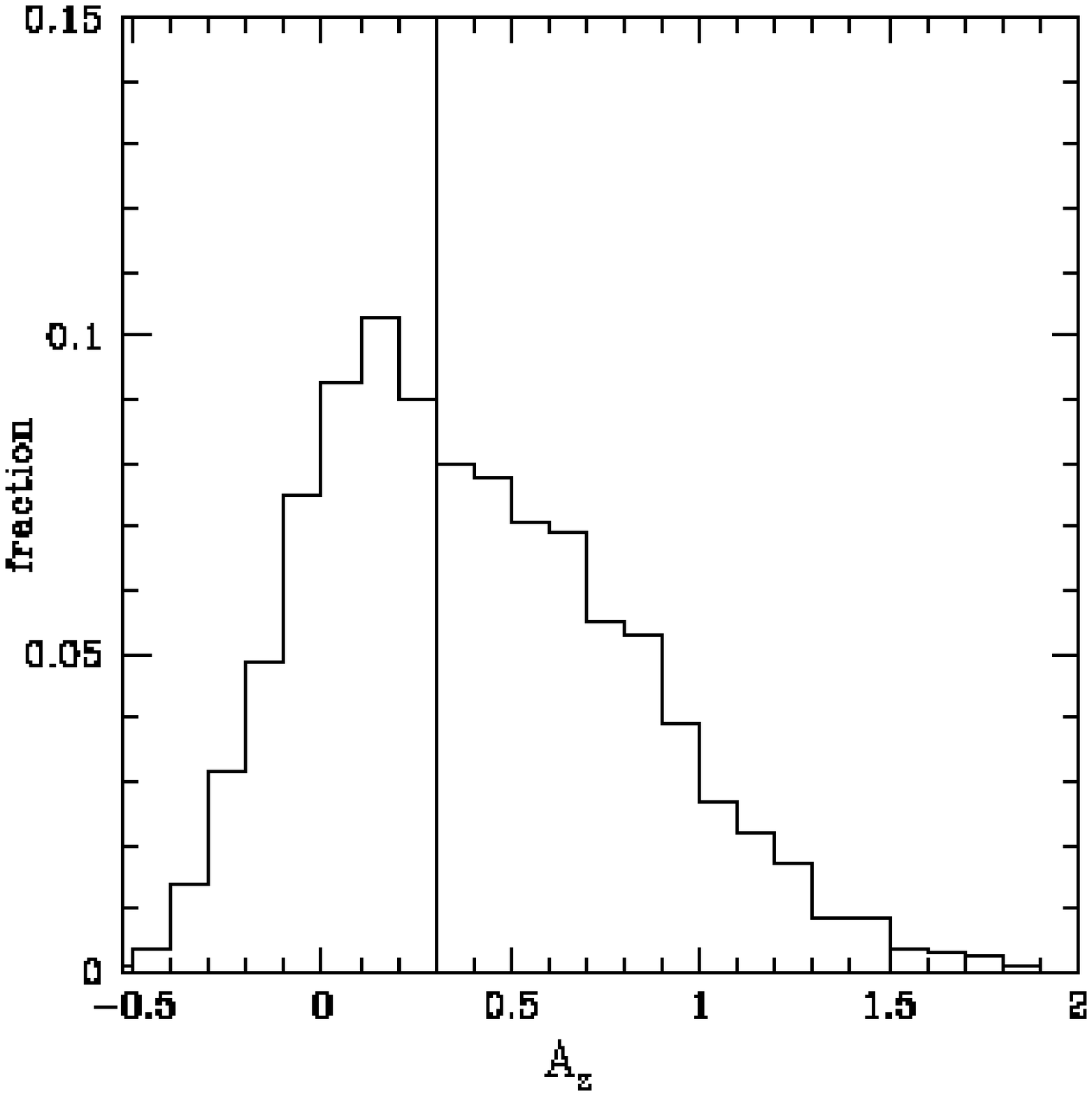}} 
}}
\centerline{\resizebox{10cm}{!}{
{\includegraphics{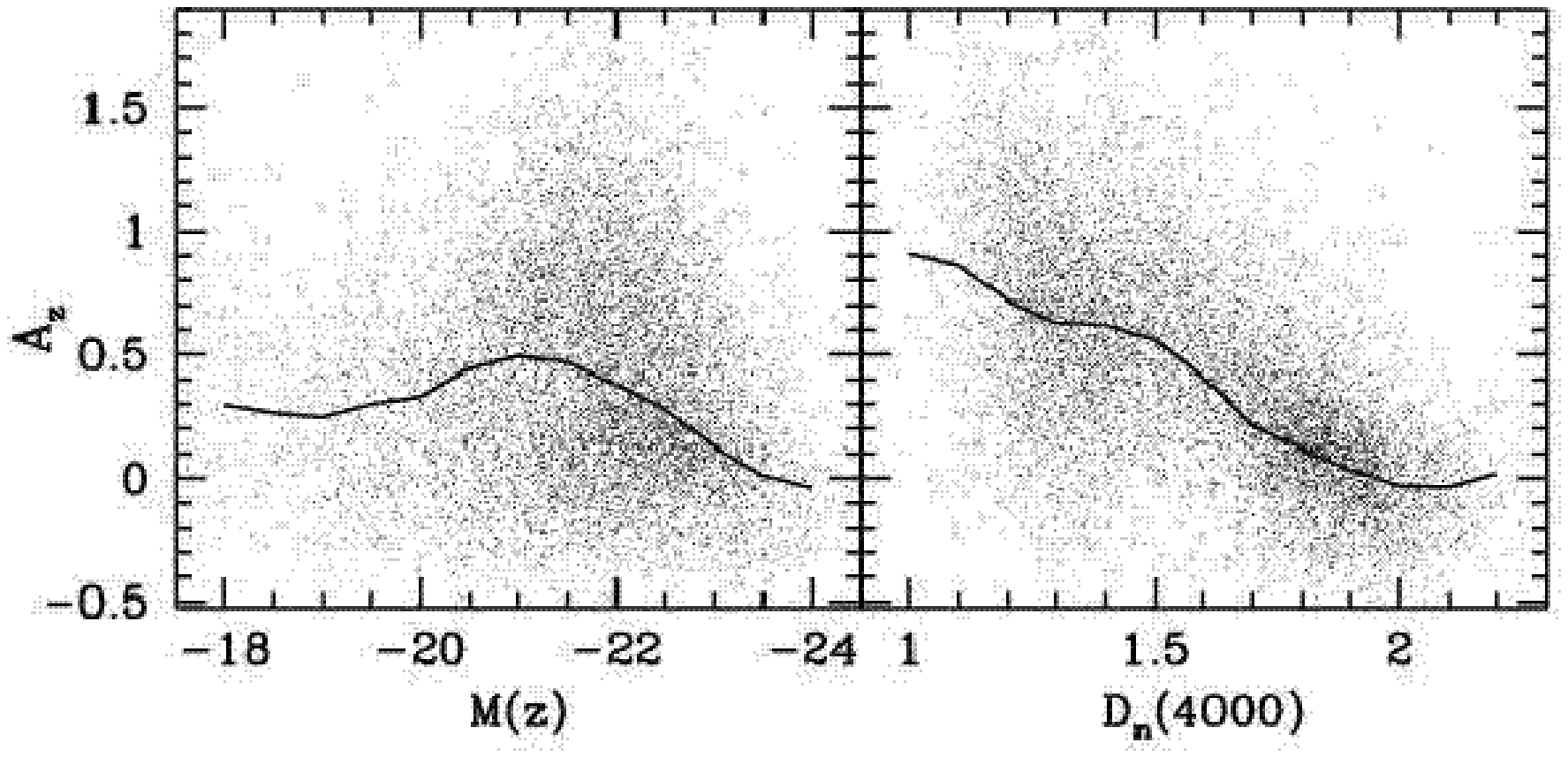}} 
}}
\caption{ {\bf Top} Histogram of the dust attenuation in the z band
(A$_z$) for the SDSS sample; 
the median value of 0.3 indicated by the vertical line, corresponds
in the B-band to 0.9 magnitude of extinction;
{\bf Bottom} A$_z$ versus z-band absolute magnitude (left), and versus 
D$_n$(4000), which is an indicator of the age of stellar population
(from Kauffmann et al. 2003a). Extinction is larger for young and small galaxies.}
\label{fig4}
\end{figure}
The distribution of dust attenuation (cf Fig. 4) reveals 
a median value of A$_z$ =0.3, which corresponds in the B-band
to  A$_B \sim$ 0.6. There is as expected a broad wing
at high extinction.  Fig. 4 shows also that the extinction
is higher for young and low-mass galaxies (the index D$_n$(4000)
being an age index).
\\
One of the most striking result of the survey is the bimodality
in the distribution of galaxies, with a separation in mass
of 3 10$^{10}$ M$_\odot$. The surface density is a function of mass,
as shown in Fig. 5. High-mass galaxies have also a high surface
density, which is about constant (HSB). The low mass galaxies
are characterized by low surface brightness (LSB), which is a power-law
function of mass (with a slope near 0.54).
The dependence of surface density with mass might well be interpreted
as a lower efficiency of star formation at the low-mass end, due to 
supernovae feedback (Dekel \& Woo 2003).

LSB galaxies have younger stellar populations, low concentrations,
and appear less evolved. The star formation history is more related
to the stellar surface density than to the stellar mass.
In the low-mass range, 50\% of galaxies are likely to have recent
bursts (10\% with large certainty), while it is only 4\% for high-mass
galaxies.
Today, starbursts (and SNe?) mostly occur in dwarfs and LSB
galaxies. It might not have been the case in the past, at high z
(Kauffmann et al. 2003b)
\begin{figure}
\centerline{\resizebox{7cm}{!}{
{\includegraphics{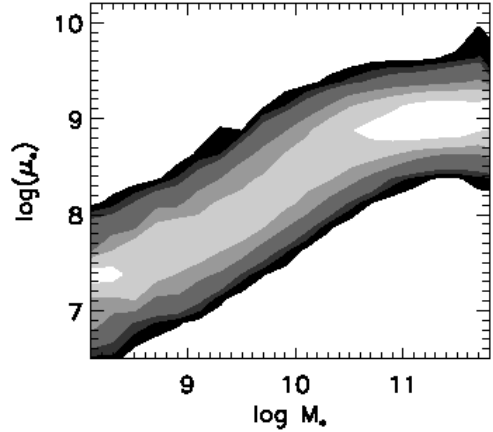}} 
}
}
\caption{ Distribution of stellar surface densities ($\mu_*$), as a function 
of stellar mass (M$_*$) for 10$^5$ galaxies in the SDSS, 
from Kauffmann et al. (2003b).}
\label{fig5}
\end{figure}
\section{Conclusions}
SNe Ia are not exactly ``standard candles'', but the shape of their
light curve allows their calibration towards this goal. Age and metallicity
effects can then be corrected, and their evolution with redshift is
being tested.
There is a surprisingly low average obscuration for high-z SN Ia, A$_B$= 0.3 mag,
twice lower than the average for galaxies, but this could be due to selection
effects.
\\
SNe Ia occur more frequently in spiral galaxies, and in outer parts of E/SO galaxies.
The scatter in their peak luminosities and in the correction factors
is larger for late-types, which are likely to be more frequent at high redshift.
This trend is however compensated by the smaller range in progenitor age,
leading to less scatter at high z.
The physics of SNe Ia explosions is not yet well known, but statistics as a function
of metallicity and age of the host stellar populations should
help to understand better the evolution of their properties.
%

\end{document}